\documentclass[a4paper,11pt]{article}
\pdfoutput=1 

\usepackage{jcappub} 

\usepackage{graphicx}
\usepackage{amsmath}
  
\newcommand\etal{{\it {et al. }}}

\newcommand{\cl}{\text{CL }}

\newcommand{\loc}{\text{loc}}

\newcommand{\BR}{BR}

\newcommand{\brbsgamma}{\BR(\overline{B}\rightarrow X_s\gamma)}

\newcommand\brbsmumu{\BR(\overline{B}_s\to\mu^+\mu^-)}

\newcommand\RBtaunu{\frac{\BR(B_u \to \tau \nu)}{\BR(B_u \to \tau \nu)_{SM}}}
\newcommand\DeltaO{\Delta_{0-}}
\newcommand\RBDtaunuBDenu{\frac{\BR(B \to D \tau \nu)}{\BR(B \to D e \nu)}}
\newcommand\Rl{R_{l23}}

\newcommand\Dstaunu{\BR(D_s \to \tau \nu)}
\newcommand\Dsmunu{\BR(D_s\to \mu \nu)} 
\newcommand\Dmunu{\BR(D \to \mu \nu)}

\newcommand{\mev}{\mbox{ MeV}}

\newcommand{\tev}{\mbox{ TeV}}

\newcommand{\mhl}{m_h}

\newcommand{\params}{{\mathbf \Theta}}

\newcommand{\like}{{\mathcal L}}

\newcommand{\vs}{v_\odot}

\newcommand{\rholoc}{\rho_\text{loc}}
\newcommand{\sigmaSI}{\sigma^\text{SI}_{\chi N}}
\newcommand{\sigmaSD}{\sigma^\text{SD}_{\chi N}}
\newcommand{\neut}{{\tilde{\chi}^0_1}}
\newcommand{\fiveinvfb}{5.8 fb$^{-1}$}

\title{\boldmath Impact of nucleon matrix element uncertainties on the interpretation of direct and indirect dark matter search results}

\author[1,3]{R. Ruiz de Austri}
\author[2,3]{C. P\'erez de los Heros}

\affiliation[1]{Instituto de F\'isica Corpuscular, IFIC-UV/CSIC, 
        Valencia, Spain}
\affiliation[2]{Department of Physics and Astronomy, 
        Uppsala University. Uppsala, Sweden}
\affiliation[3]{Kavli Institute for Theoretical Physics, University of California, Santa Barbara, USA}

\emailAdd{rruiz@ific.uv.es}
\emailAdd{cph@physics.uu.se}

\abstract{ 
 We study in detail the impact of the current uncertainty in nucleon matrix elements 
on the sensitivity of direct and indirect experimental techniques for dark matter detection. 
We perform two scans in the framework of the cMSSM: one using recent values of the pion-sigma 
term obtained from Lattice QCD, and the other using values derived from experimental 
measurements. The two choices correspond to extreme values quoted in the literature 
and reflect the current tension between different ways of obtaining information about the structure 
of the nucleon. All other inputs in the scans, astrophysical and from particle physics, are kept unchanged. 
We use two experiments, XENON100 and IceCube, as benchmark cases to illustrate our case. 
We find that the interpretation of dark matter search results from direct detection experiments is 
more sensitive to the choice of the central values of the hadronic inputs than the results of 
indirect search experiments. The allowed regions of cMSSM parameter space after including XENON100 
constraints strongly differ depending on the assumptions on the hadronic matrix elements used. On the 
other hand, the constraining potential of IceCube is almost independent of the choice of these values.
}


\begin{document}
\maketitle
\flushbottom

\section{Introduction} \label{sec:intro}

 The search for dark matter is one of the most active fields in particle and astroparticle physics today. 
Generic candidates that can account for the cold dark matter required to fit cosmological observations must be 
stable, massive and weakly interacting with normal matter, and they are usually referred to as WIMPs (weakly 
interacting massive particles). There are various theoretical possibilities providing viable candidates for dark 
matter. Natural choices arise in extensions of the Standard Model of particle physics, where new particles are 
introduced which have the right lifetime and annihilation cross section to have survived as thermal relics from 
the early universe. In particular, stable particles predicted in different flavors of the Minimal Supersymmetric 
extension to the Standard Model (MSSM) have been extensively studied as dark matter candidates, all having in 
common that they have weak-type cross sections with ordinary matter. A good dark matter relic particle is the 
lightest neutralino, which is the mass eigenstate of a mixture of bino, wino (the superpartners of the $B$, $W^0$ 
gauge bosons) and higgsinos (the superpartners of the $H^0_1$, $H^0_2$ Higgs bosons). In a wide class of models, 
the neutralino is the lightest stable supersymmetric particle. \par

  A realistic interpretation of the results of dark  matter searches must include not only the systematic 
uncertainties of the experiments themselves, but also uncertainties in the ingredients which enter 
in the calculation of expected signals. These include uncertainties in astrophysical inputs as well as in 
nuclear physics inputs. Recent parameter scans of different supersymmetric scenarios, including the present work, 
include uncertainties in the estimation of the local dark matter density and velocity dispersion, as well as  
in nuclear physics quantities, like in the hadronic nucleon matrix elements
~\cite{Bertone:2011nj, Strege:2011pk, Buchmueller:2012hv, Strege:2012bt, Fowlie:2013oua}. 
However, it was already pointed out in 1996 by the authors in~\cite{Jungman:1995df} that 
``the uncertainty in the pion-nucleon sigma term is perhaps the largest source of uncertainty'' 
in the calculation of the WIMP-nucleon cross section, and therefore on the interpretation of the 
results from dark matter search experiments. The authors of~\cite{Bottino:1999ei, Bottino:2001dj, Ellis:2008hf} 
have also brought up the issue of the quark content of the nucleon in the context of the WIMP-nucleon cross section, 
specifically on the strange-quark component. Indeed the capture of WIMPs in celestial bodies and their 
scattering off target nuclei in direct detection experiments, depend on the WIMP-nucleus cross 
section, the results from both techniques being complementary~\cite{Alina:13a}. But the
 calculation of the WIMP-nucleus cross section from the fundamental interactions of WIMPs with quarks and gluons involves 
two further layers of complexity: the parametrization of the nucleon structure in terms of quark and gluon structure 
functions and the description of a nuclear state as a coherent superposition of nucleons~\cite{Engel:92a}.
Already the first step presents experimental challenges since the structure of nucleons at low momentum transfer is 
difficult to directly probe experimentally, specially the strangeness content. Recent progress from  
Lattice QCD in combination with 
chiral effective field theory, has improved considerably the uncertainties in the calculations of the hadron masses 
and the light quark and strange quark sigma terms~\cite{Young:09a,Bali:12a}. But there is still some tension with 
the central values obtained from experimental results. The second step, the description of a nucleus as 
a system of nucleons, is relevant when considering different target nuclei for capture or scattering, where  
the uncertainties in the nuclear from factors can play an important role in the calculations of 
the total WIMP-nucleus cross section. A recent analysis in the case of direct searches can be found in~\cite{Cerdeno:13a}.

 In this paper we focus on the effect of using different estimations of the strangeness component 
of the nucleon on the interpretation of signal rates in neutrino telescopes, and its correlation with 
the results from direct search experiments.

We extend the method developed in~\cite{Trotta:09a} and we work in the context of the 
Constrained Minimal Supersymmetric Extension of the Standard Model 
(cMSSM)~\cite{Kane:1993td}. We will assume that the lightest neutralino is the dark matter candidate. 
We use the effective area and angular resolution of IceCube in its 86-string configuration to make concrete predictions 
on the effect on the cMSSM, based on their null result on dark matter search from the Sun, but the conclusions can 
easily be extended to any generic neutrino telescope. For direct searches we use the most recent XENON100 results.  
Keeping all other inputs unchanged, we perform scans over the cMSSM parameter space using different values for the 
hadronic variables involved.  
In a first scan, we use the value of the pion-nucleon sigma term, $\sigma_{\pi N}$, and the
contribution of strangeness to the total proton spin from recent lattice QCD estimations.
We label results from this choice as "LQCD" in the rest of the paper.
In a second approach, we base our inputs for the nucleon matrix elements on experimentally
obtained values of the spin content of the nucleon and of $\sigma_{\pi N}$.
Given that there is a wide range of values in the literature for $\sigma_{\pi N}$, extracted
from accelerator data using different methods, we chose a high value of 74$\pm$12~MeV
which strongly differs from the LQCD value of 43$\pm$6.1~MeV. We cover in this way the
range of values of $\sigma_{\pi N}$ quoted by different groups, and we illustrate the
maximum effect that the uncertainty on the determination of this quantity can have on dark
matter searches. We label results from this second choice as "Experiment" in the rest of the paper.
In all cases, we compute the expected number of events in IceCube and XENON100 and evaluate the power of these 
experiments to constrain the cMSSM parameter space, the differences in each case being due to the different 
choices of inputs for the nucleon matrix elements. We note that even if we work in the framework of the cMSSM, 
the effect of the different choices of the  matrix elements are rather decoupled from the details of the model, 
since they enter as a scaling factor in the calculation of the WIMP-nucleon cross section (see equations~\ref{eq:f_t} and 
\ref{eq:a_n} below). So a generalization of our conclusions to other generic supersymmetric models is possible.\par

 The paper is organized as follows. In section~\ref{sec:Xsection} we summarize the relevant components of 
the neutralino cross section with nucleons.
In section~\ref{sec:framework} we present the theoretical and statistical framework 
and provide details about the supersymmetric model we study, the nuisance parameters, 
the implementation of the experimental constraints as well as the scanning technique used. 
In section \ref{sec:results} we present the results from the different scans and comment on the 
conclusions that can be extracted for direct and indirect dark matter searches. Finally, 
we present our conclusions in section \ref{sec:concl}. 

\section{The WIMP-nucleon cross section and nucleon matrix elements}\label{sec:Xsection}

 The interaction of WIMPs with a target material depends on a first instance on the total scattering cross section 
of the WIMP with the different nuclear species that make up the target. The WIMP-nucleus cross section can in turn be 
parametrized as a function of the fundamental WIMP-nucleon scattering cross section, which can be 
decomposed into a spin-independent and an spin-dependent part~\cite{Jungman:1995df}. The spin-independent component, 
$\sigma^{SI}_{\chi N}$, is proportional to the square of the effective coupling of the neutralino to the nucleon, $f_N$
\begin{equation}
\frac{f_N}{m_N} = \sum_{q=u,d,s} f_{Tq}^{N} \frac{\alpha_{q}^s}{m_q} + \frac{2}{27} f_{TQ}^{N} \sum_{q=u,d,s} \frac{\alpha_{q}^s}{m_q},
\label{eq:f_t}
\end{equation}
where N stands for p or n. The coefficients $\alpha_{q}^s$ refer to the neutralino-quark scalar couplings in the 
low-energy effective Lagrangian and are calculated for each point in our scans. 
The coefficients $f_{Tq}^{N}$ are the nucleon matrix elements and represent the contributions of the light quarks 
to the mass of the nucleon. The second term corresponds to the interaction of the neutralino with the gluon scalar 
density in the nucleon, with $f_{TQ}^{N}=1-\sum_{q=u,d,s}f_{Tq}^{N}$. The matrix elements are defined as 
$f_{Tq}^{N}=\frac{m_q}{m_N}  \langle N | \bar{q}q | N \rangle$ and can be calculated in Lattice QCD (LQCD), or derived 
experimentally from measurements of the pion-nucleon sigma term (they are essentially proportional to  
$\sigma_{\pi N}$ through a proportionality constant which is a function of the quark masses~\cite{Ellis:2008hf}). 
The pion-nucleon sigma term can be extracted from $\pi-$N scattering experiments and the results are extrapolated to 
zero exchange momentum using chiral perturbation theory (ChPT). Still, both the experimental measurements and the LQCD 
calculations of $\sigma_{\pi N}$ are plagued with difficulties and systematic uncertainties which result in quoted values in 
the literature between about 40~MeV and 80~MeV, with errors between 10\% and 20\% on these numbers. 
The range of these values (especially that derived for $f_{Ts}^{N}$) has an important impact on the computed 
spin-independent cross section, and therefore in the interpretation of the constraints from direct experiments which use 
spin-0 target nuclei. 
 Lattice calculations of $f_{Ts}^{N}$ tend to agree with the lower values extracted from experimental analyses. The 
Lattice world average gives a value of 0.043$\pm$0.011 for $f_{Ts}^{N}$~\cite{Junnarkar:13a}, while one obtains 
$f_{Ts}^{N}=0.046 \pm 0.013$ from recent data from the CHAOS spectrometer~\cite{Stahov:12a}. But analyses of the mass 
spectrum of exotic baryons~\cite{Schweitzer:04a} or a partial wave analysis of TRIUMF $\pi$N scattering 
data~\cite{Pavan:02a}, provide much higher values for $f_{Ts}^{N}$, of about 0.4. We have performed scans using two 
extreme values of $f_{Ts}^{N}$ (0.043 and 0.493) in order to illustrate the dependence of the interpretation of experimental 
dark matter searches on the current lack of precise knowledge of the value of $f_{Ts}^{N}$. In both cases we include the 
associated uncertainty on the central values as a nuisance parameter. \par

 The spin-dependent part of the  WIMP-nucleon cross section, $\sigma^{SD}_{\chi N}$, is of relevance for indirect 
experiments searching for dark matter accumulated in the Sun. $\sigma^{SD}_{\chi N}$ is proportional to the square 
of $a_p \left < S_p \right > + a_n \left < S_n \right >$, where $\left < S_{p/n} \right >$ is the expectation value of 
the spin content of the protons/neutrons in the nucleus, and the factors $a_{p/n}$ are defined as
\begin{equation}
a_N = \sum_{q=u,d,s} \frac{\alpha^a_{q}}{\sqrt{2} G_f} \Delta_q^N, 
\label{eq:a_n}
\end{equation}
where, again, N stands for p or n. The axial-vector matrix elements $\Delta_q^N$ contain information about the quark spin 
content of the nucleon and are proportional to $\langle N | \bar{q} \gamma_{\mu} \gamma_{5} q | N \rangle$.
The coefficients $\alpha^a_{q}$ are the fundamental neutralino-quark axial couplings in the effective Lagrangian 
of the model. The factors $\Delta_q^N$ are better known than the $f_{Tq}^{N}$s. Indeed the LQCD calculations~\cite{Bali:12a} and 
the experimental values obtained by COMPASS~\cite{COMPASS:10a,COMPASS:08a} for the $\Delta_u^N$ and $\Delta_d^N$ agree 
within 10\% (see table~\ref{tab:nuis_params}). The COMPASS results are in agreement also with previous results from 
HERMES and SMC~\cite{Aidala:13a}. There is some tension, on the other hand, in the value of $\Delta_s^N$ obtained from 
LQCD and the experimental measurement. Even if $\Delta_s^N$ is an order of magnitude smaller than $\Delta_u^N$ or 
$\Delta_d^N$, and its contribution is sub-dominant to the total cross section, we have performed scans using the value 
obtained from LQCD and from the COMPASS measurement. This covers the current range of uncertainty on $\Delta_s^N$  
and allows us to evaluate its impact on the interpretation of indirect dark matter searches.\par


\section{Theoretical and statistical framework} \label{sec:framework}

 We work in the framework of the cMSSM, which assumes that supersymmetry is broken softly by gravity mediation~\cite{Kane:1993td}
and the soft-parameters are universal at a high scale ($M_X$). Hence the model can be parametrized in terms of four free 
parameters: the common scalar mass, $m_0$, the gaugino mass $m_{1/2}$, 
the coefficient of the trilinear interaction $A_0$ plus the ratio between the vacuum expectation values of the Higgs bosons $\tan\beta$. 
Additionally the sign of the higgsino mass parameter, $\mu$, needs to be fixed. 
The value of $\mu$ is determined from the conditions of radiative electroweak symmetry breaking.  We fix $\text{sign}(\mu)$=~+~1, 
motivated by consistency arguments involving measurements of the anomalous muon magnetic moment. 

We use Bayesian methods for doing inference of the cMSSM, the key ingredient being  Bayes's
theorem, namely 
\begin{equation}
p(\params|\mathbf{D})=\frac{p(\mathbf{D}|\params) p(\params)}{p(\mathbf{D}}, 
\end{equation}
where  $\mathbf{D}$ are the data and $\params$ are the model parameters of interest. 
The equation reflects the fact that the posterior probability 
distribution function (pdf) $p(\params|\mathbf{D})$ for the parameters is obtained from the likelihood function 
$p(\mathbf{D}|\params) \equiv \like(\params)$ and the prior pdf $p(\params)$. The Bayesian evidence $p(\mathbf{D})$ is 
a normalization constant which in the case of model comparison can be ignored. 

In order to study the constraints on a single parameter of interest $\theta_i$, one considers the one-dimensional 
marginal posterior pdf. The marginal pdf is obtained from the full posterior distribution by integrating (marginalising) over 
the unwanted parameters in the $n$-dimensional parameter space
\begin{equation}
p(\theta_i|\mathbf{D})=\int p(\params|\mathbf{D}) d\theta_1 ... d\theta_{i-1} d\theta_{i+1} ... d\theta_{n}.
\end{equation}

Since we are only interested in studying the effect of different hadronic inputs in the calculations, and not   
to assess the degree of dependency of our results on the choice of priors, we have adopted ``log'' priors in $m_0, m_{1/2}$ 
and a ``flat" prior on $A_0$ and $\tan\beta$. 
The reason is that the soft breaking masses can take any value between, say, the electro-weak scale and a few TeV 
(to avoid fine-tunning) with the same a priori weight.  This is achieved assuming a flat prior on the log of those 
parameters. 
For a discussion on the dependency of Bayesian model parameter scans on the choice of prior see 
for example~\cite{Trotta:2008bp}. The ranges scanned for the cMSSM parameters are 
$50 \mev \leq m_0, m_{1/2} \leq 8\tev $, $-7 \tev \leq A_0 \leq 7 \tev$ and $2 \leq \tan\beta \leq 62$
 \footnote{Our motivation for this choice is based on the naturalness criterium which is connected 
to the fact that SUSY masses above  a few TeV lead to a large fine-tunning to reproduce the electro-weak 
scale. In this respect we are conservative.}

In addition to the cMSSM model parameters, we include three categories of nuisance parameters in the analysis. 
Those accounting for the uncertainties on measurements in some of the Standard Model parameters which 
have been shown to have an important impact in inferences of SUSY models~\cite{Roszkowski:2007fd}, 
and those from astrophysics and nuclear physics  which enter at the level of dark matter direct and indirect detection 
constraints. As astrophysical nuisance parameters we consider the local dark matter density $\rho_{\loc}$, and 
two quantities parameterizing the local WIMP velocity distribution: the velocity of the Sun in the Galaxy, $\vs$, and 
the velocity dispersion of WIMPs in the halo,$v_d$, assumed Maxwellian~\cite{DarkSusy}. For the hadronic nuisances  
we include the hadronic nucleon matrix elements for both spin-independent ($f_{Tu}$, $f_{Td}$ and $f_{Ts}$) and spin-dependent 
($\Delta_u$, $\Delta_d$ and $\Delta_s$) WIMP--nucleon cross sections. For them we adopt informative Gaussian priors
as mentioned in Section~\ref{sec:Xsection}. Table~\ref{tab:nuis_params} summarizes the values used for the different 
nuisance parameters considered.

\subsection{Experimental constraints}
 
\begin{table}
\begin{center}
\begin{tabular}{|l l l l|}
\hline
\hline
\multicolumn{4}{|c|}{Nuisance parameters} \\
\hline
\multicolumn{4}{|c|}{Standard Model} \\
\hline
$M_t$ [GeV] & $173.1 \pm 1.3$  &   & \cite{pdg12} \\
$m_b(m_b)^{\bar{MS}}$ [GeV] & $4.20\pm 0.07$ &   & \cite{pdg12}\\
$[\alpha_{em}(M_Z)^{\bar{MS}}]^{-1}$ & $127.955 \pm 0.030$ &   & \cite{pdg12}\\
$\alpha_s(M_Z)^{\bar{MS}}$ & $0.1176 \pm  0.0020$ &    & \cite{Hagiwara:2006jt}\\
\hline
\multicolumn{4}{|c|}{Astrophysical} \\
\hline
$\rho_{\loc}$ [GeV/cm$^3$] & $0.4\pm 0.1$ & &  \cite{Pato:2010zk}\\
$\vs$ [km/s] & $230.0 \pm 30.0$ & & \cite{Pato:2010zk}\\
$v_d$ [km/s] & $282.0 \pm 37.0$ &  & \cite{Pato:2010zk}\\
\hline
\multicolumn{4}{|c|}{Hadronic} \\
& \multicolumn{2}{c}{LQCD \hfill Experiment} &\\
\hline
$f_{Tu}$ & $ 0.0190\pm 0.0029$ &  $ 0.0308 \pm 0.0061$ & \cite {Ren:12a}, \cite{Schweitzer:04a}\\
$f_{Td}$ & $ 0.0246\pm 0.0037$ & $0.0459 \pm 0.0089$ & \cite {Ren:12a}, \cite{Schweitzer:04a}\\
$f_{Ts}$ & $ 0.043  \pm 0.011$ & $0.493\pm 0.159$ & \cite {Junnarkar:13a}, \cite{Schweitzer:04a}\\
$\Delta_u$ & $ 0.787 \pm 0.158 $ & $0.75 \pm 0.05$ & \cite{Bali:12a}, \cite{COMPASS:10a}  \\
$\Delta_d$ & $ -0.319 \pm 0.066 $ & $-0.34 \pm 0.07$ & \cite{Bali:12a}, \cite{COMPASS:10a}  \\
$\Delta_s$ & $ -0.020 \pm 0.011 $ & $-0.09 \pm 0.02$ & \cite{Bali:12a}, \cite{COMPASS:08a} \\
\hline
\end{tabular}
\end{center}
\caption{\fontsize{9}{9}\selectfont Nuisance parameters adopted in the scans of the cMSSM parameter space, indicating 
the mean and standard deviation used for the Gaussian prior on each of them. The matrix elements $f_{Tu}$ and 
 $f_{Td}$ are extracted from the value of $\sigma_{\pi N}$ following~\cite{Ellis:2008hf}. We use values 
of $\sigma_{\pi N}$ derived from LQCD calculations ($\sigma_{\pi N}$=43$\pm$6.1~MeV~\cite {Ren:12a}) and 
derived from the mass spectrum of exotic baryons ($\sigma_{\pi N}$=74$\pm$12 MeV~\cite{Schweitzer:04a}) as two 
extreme representative values of the range of $\sigma_{\pi N}$ found in the literature. For the LQCD value of $f_{Ts}$ 
we use the most recent world average from LQCD calculations, $\sigma_{\pi N}$=40$\pm$10~MeV~\cite {Junnarkar:13a}. 
}
\label{tab:nuis_params}
\end{table}

The likelihood function is composed of several different parts, corresponding to the different experimental constraints that are 
applied in our analysis:
\begin{equation}
\ln \like = \ln \like_\text{LHC} + \ln \like_\text{Planck}  + \ln \like_\text{EW}+ \ln \like_\text{B(D)} + \ln \like_{g-2}+ \ln \like_\text{Xe100} + \ln \like_\text{IC86} .      
\end{equation}

The LHC likelihood implements recent null results from SUSY searches from ATLAS. Exclusion limits in the ($m_0$,$m_{1/2}$) 
plane are based on a search by the ATLAS collaboration for squarks and gluinos in final states that contain missing $E_T$, 
jets and 0 leptons in 5.8 fb $^{-1}$ integrated luminosity of data at $\sqrt{s} = 8$ TeV collision energy \cite{LHCSUSY}.  
The LHC exclusion limit is included in the likelihood function by defining the likelihood of samples corresponding to masses 
below the limit to be zero.
We furthermore include the most recent experimental constraint from the CMS and ATLAS collaborations on the mass of the lightest 
Higgs boson which combination is $m_h = 125.66 \pm 0.41$ GeV \cite{Higgs}. We use a Gaussian likelihood and we add in quadrature 
a theoretical error of 2 GeV to the experimental error.
We also include the new LHCb constraint on $\brbsmumu = (3.2^{+1.5}_{-1.2}) \times 10^{-9}$, derived from a combined analysis 
of 1 fb$^{-1}$ data at $\sqrt{s} = 7$ TeV collision energy and 1.1 fb$^{-1}$ data at $\sqrt{s} = 8$ TeV collision energy \cite{Aaij:2012ct}. 
We implement this constraint as a Gaussian distribution with a conservative experimental error of $\sigma = 1.5 \times 10^{-9}$, 
and a $10\%$ theoretical error. 

The constraint from the dark matter relic abundance is included as a Gaussian in $\ln \like_\text{Planck}$. We use the recent 
PLANCK value $\Omega_\chi h^2 = 0.1196 \pm 0.0031$~\cite{Planck:13a} and we add a fixed 10\% theoretical uncertainty in quadrature. 
We assume that neutralinos make up all of the dark matter in the universe. 

$\ln \like_\text{EW}$ implements precision tests of the electroweak sector. The electroweak precision observables
$M_W$ and $\sin^2\theta_{eff}$ are included with a Gaussian likelihood. 

Relevant constraints from $B$ and $D$ physics are included in $\ln \like_\text{B(D)}$ as a Gaussian likelihood. The full 
list of $B$ and $D$ physics observables included in our analysis is shown in table~\ref{tab:exp_constraints}.

The measured anomalous magnetic moment of the muon, included as a Gaussian datum in $\ln \like_{g-2}$, provides important 
information about the supersymmetric parameter space, since it can be experimentally measured to very good precision. 
By comparing the theoretical value of this quantity favored in the Standard Model with the experimental result the 
supersymmetric contribution $\delta a_{\mu}^{SUSY}$ can be constrained. The experimental measurement of the muon anomalous 
magnetic moment based on e$^+$e$^-$ data is in tension with the Standard Model prediction by 
$\delta a_{\mu}^{SUSY} = (28.7 \pm 8.0) \times 10^{-9}$ \cite{Davier:2010nc}, i.e., a $3.6\sigma$ 
discrepancy between the experimental result and the expected 
value from Standard Model physics alone. 

As for constraints from direct dark matter search experiments we use the recent results from XENON100
with 225 live days of data collected  between February 2011 and March 2012 with 34 kg fiducial volume~\cite{XENON100}. 
We calculate the number of expected signal recoil events from each of the points on our scan as
\begin{equation}
  N_s^{XE}(\params) \,=\, T \frac{\rholoc} {m_{\chi}}\frac{M}{m_A } \cdot  \int_{E_{\text{thr}}} dE\, 
\int_{v<v_{\text{esc}}} d\text{v}\, v f(\text{v}+\text{v}_{\text{Earth}})\frac{d\sigma_{\chi N}(\params,v,E)}{dE}\,\, ,    
\label{eq:n_events02}
\end{equation}
where T and M are the exposure of and total mass of the detector respectively, $m_A$ is the detector nucleus target mass, 
$\rholoc$ is the local dark matter density, $\text{v}$ the dark matter velocity 
in the detector rest frame, $\text{v}_{\text{Earth}}$ the Earth velocity in the rest frame of the Galaxy, $f(.)$ the dark 
matter velocity distribution function (assumed Maxwellian) and $d\sigma_{\chi N}/dE$ is the differential cross section for 
the interaction between the neutralino and the nucleus. The integral is over the volume  of phase space for which $v$ 
is smaller than the escape velocity $v_{\text{esc}}$ and larger than the minimal velocity $v_{\text{min}}(E)$ able to produce
a recoil with energy E, above the detector energy threshold, $E_{\text{thr}}$. 
We follow the treatment of XENON100 data as described in detail in~\cite{Strege:2012bt}, building 
the likelihood function, $\ln \like_\text{Xe100}$, as a Poisson distribution for observing N recoil events when  
$N_s(\params)$ signal plus $N_b$ background events are expected. The expected number of events from the background-only 
hypothesis in the XENON100 run is $N_b$ = 1.0 $\pm$ 0.2, while  the collaboration reported N=2 events observed in the 
pre-defined signal region. 
We use the latest values for the fiducial mass and exposure time of the detector, and we include the reduction of the lower energy 
threshold for the analysis to 3 photoelectron events 
 and an update to the response to 122 keV gamma-rays to $2.28$ PE/keVee, obtained from new calibration measurements, in 
accordance with the values reported in Ref.~\cite{XENON100}. We make the simplifying assumption of an 
energy-independent acceptance of data quality cuts, and adjust the acceptance-corrected exposure to accurately reproduce 
the exclusion limit in the $(m_{\neut},\sigmaSI)$ plane reported in Ref.~\cite{XENON100} in the mass range of 
interest.

In a similar fashion, the likelihood of IceCube, $\ln \like_\text{IC86}$, is based on the number of signal events 
expected, 
\begin{equation}
  N_s^{IC}(\params) \,=\, T \cdot \int_{E_{thr}} \frac{dN_{\nu}(\params)}{dt dE_{\nu} dA}\,A_{eff}(E_{\nu})\,dE_{\nu}\, ,    
\label{eq:n_events01}
\end{equation}
where $T$ is the exposure time, $A_{eff}$ is the detector effective area,  E$_{thr}$ is the energy threshold of the 
detector and $dN_{\nu}/dE_{\nu}dA$ is the differential muon--neutrino flux at the Earth from WIMP annihilation for a 
given choice of cMSSM parameters. The effective area is a measure of the efficiency of the detector to the signal, 
and includes the neutrino-nucleon interaction probability, the energy loss of the produced muon from the interaction 
point to the detector and the detector trigger and analysis efficiency.  We use the public information about the 
86-string configuration of IceCube released with DarkSUSY, ie, the estimated effective area of the detector and the 
background pdf. IceCube has reported no signal on their searches for an excess neutrino flux from the Sun in their analysis 
with the 79-string configuration ~\cite{IceCube:13a}, so we assume here a background--only scenario. 
We build the Likelihood for IceCube following~\cite{Scott:12a}: given a number of signal events for a given model and the 
number of estimated background events obtained by sampling the background pdf, we use a Poisson likelihood convoluted 
with a log-normal distribution for the uncertainty on the estimation of the number of signal events. 
This uncertainty takes into account the uncertainty in the effective area and in the signal prediction for which we use a 
10\% relative error for this quantity. We further use an angular cut of $\phi_{\text cut} = 10^{\text o}$ around the 
solar position.  
We normalize our results to one calendar year of IceCube data taking. 
We assume no contamination in the background estimation from remaining misreconstructed atmospheric muons. 
We have neglected the potential contribution to the background from cosmic ray interactions in the Sun corona. 
This flux has been estimated in~\cite{Fogli:06a} and predicts 
about 1 event per year in IceCube. We have also neglected the 
effects arising from uncertainties in the Solar composition~\cite{Ellis:2009ka} and in the capture rate from other 
planets, which has been shown to be negligible~\cite{Sivertsson:12a}.

\begin{table*}
\begin{center}
\begin{tabular}{|l | l l l | l|}
\hline
\hline
Observable & Mean value & \multicolumn{2}{c|}{Uncertainties} & Ref. \\
 &   $\mu$      & ${\sigma}$ (exper.)  & $\tau$ (theor.) & \\\hline
$M_W$ [GeV] & 80.399 & 0.023 & 0.015 & \cite{lepwwg} \\
$\sin^2\theta_{eff}$ & 0.23153 & 0.00016 & 0.00015 & \cite{lepwwg} \\
$\delta a_\mu^{\mathrm{SUSY}} \times 10^{10}$ & 28.7 & 8.0 & 2.0 & \cite{Davier:2010nc} \\
$\brbsgamma \times 10^4$ & 3.55 & 0.26 & 0.30 & \cite{hfag}\\
$R_{\Delta M_{B_s}}$ & 1.04 & 0.11 & - & \cite{deltambs} \\
$\RBtaunu$   &  1.63  & 0.54  & - & \cite{hfag}  \\
$\DeltaO  \times 10^{2}$   &  3.1 & 2.3  & - & \cite{delta0}  \\
$\RBDtaunuBDenu \times 10^{2}$ & 41.6 & 12.8 & 3.5  & \cite{Aubert:2007dsa}  \\
$\Rl$ & 0.999 & 0.007 & -  &  \cite{Antonelli:2008jg}  \\
$\Dstaunu \times 10^{2}$ & 5.38 & 0.32 & 0.2  & \cite{hfag}  \\
$\Dsmunu  \times 10^{3}$ & 5.81 & 0.43 & 0.2  & \cite{hfag}  \\
$\Dmunu \times 10^{4}$  & 3.82  & 0.33 & 0.2  & \cite{hfag} \\
$\Omega_\chi h^2$ & 0.1196 & 0.0031 & 0.012 & \cite{Planck:13a} \\
$\mhl$ [GeV] & 125.66  & 0.41  & 2.0 & \cite{Higgs} \\
$\brbsmumu$ &  $3.2 \times 10^{-9}$ & $1.5  \times 10^{-9}$ & 10\% & \cite{Aaij:2012ct}\\
\hline\hline
   &  Limit (95\%~$\cl$)  & \multicolumn{2}{r|}{$\tau$ (theor.)} & Ref. \\ \hline
Sparticle masses  &  \multicolumn{3}{c|}{As in Table~4 of
  Ref.~\cite{deAustri:2006pe}.}  & \\
$m_0, m_{1/2}$ & \multicolumn{3}{l|}{ATLAS, $\sqrt{s} = 8$ TeV, \fiveinvfb 2012 limits} & \cite{LHCSUSY} \\
$m_\chi - \sigmaSI$ & \multicolumn{3}{l|}{XENON100 2012 limits ($224.6 \times 34$ kg days)} & \cite{XENON100} \\
\hline
\end{tabular}
\end{center}
\caption{\fontsize{9}{9} \selectfont Summary of the observables used
for the computation of the likelihood function
For each quantity we use a
likelihood function with mean $\mu$ and standard deviation $s =
\sqrt{\sigma^2+ \tau^2}$, where $\sigma$ is the experimental
uncertainty and $\tau$ represents our estimate of the theoretical
uncertainty. Lower part: Observables for which only limits currently
exist. The explicit form of the likelihood function is given in
ref.~\cite{deAustri:2006pe}, including in particular a smearing out of
experimental errors and limits to include an appropriate theoretical
uncertainty in the observables.
\label{tab:exp_constraints}}
\end{table*}

\subsection{Scanning technique}

The full list of experimental constraints included in the likelihood function is given in Table~\ref{tab:exp_constraints}.
In order to explore the posterior pdf we use the \texttt{SuperBayeS-v2.0} package~\cite{SuperBayes}. 
This latest version of \texttt{SuperBayeS} is interfaced with SoftSUSY 3.2.7 as SUSY spectrum calculator, 
MicrOMEGAs 2.4 \cite{MicrOMEGAs} to compute the abundance of dark matter,
DarkSUSY 5.0.5~\cite{DarkSusy} for the computation of $\sigmaSI$ and $\sigmaSD$, SuperIso 3.0 \cite{SuperIso} 
to compute $\delta a_\mu^{\mathrm{SUSY}}$ and B(D) 
physics observables, SusyBSG 1.5 for the determination of $\brbsgamma$~\cite{SusyBSG}. 

The  \texttt{SuperBayeS-v2.0} package uses the publicly available MultiNest v2.18~\cite{Feroz:2007kg,Feroz:2008xx} 
nested sampling algorithm to explore the cMSSM  model 
parameter space. MultiNest has been developed in such a way as to be an extremely efficient sampler 
even for likelihood functions defined over a parameter space of large dimensionality with a very complex structure as 
it is the case of the cMSSM. The main purpose of the Multinest is the
computation of the Bayesian evidence and its uncertainty but it produces posterior inferences as a by--product. 
Besides it is also able to reliably evaluate the profile likelihood, given appropriate MultiNest settings, 
as demonstrated in~\cite{Feroz:2011bj}. 

\section{Results} \label{sec:results}

\begin{figure}[t]
\includegraphics[angle=0,width=.48\textwidth]{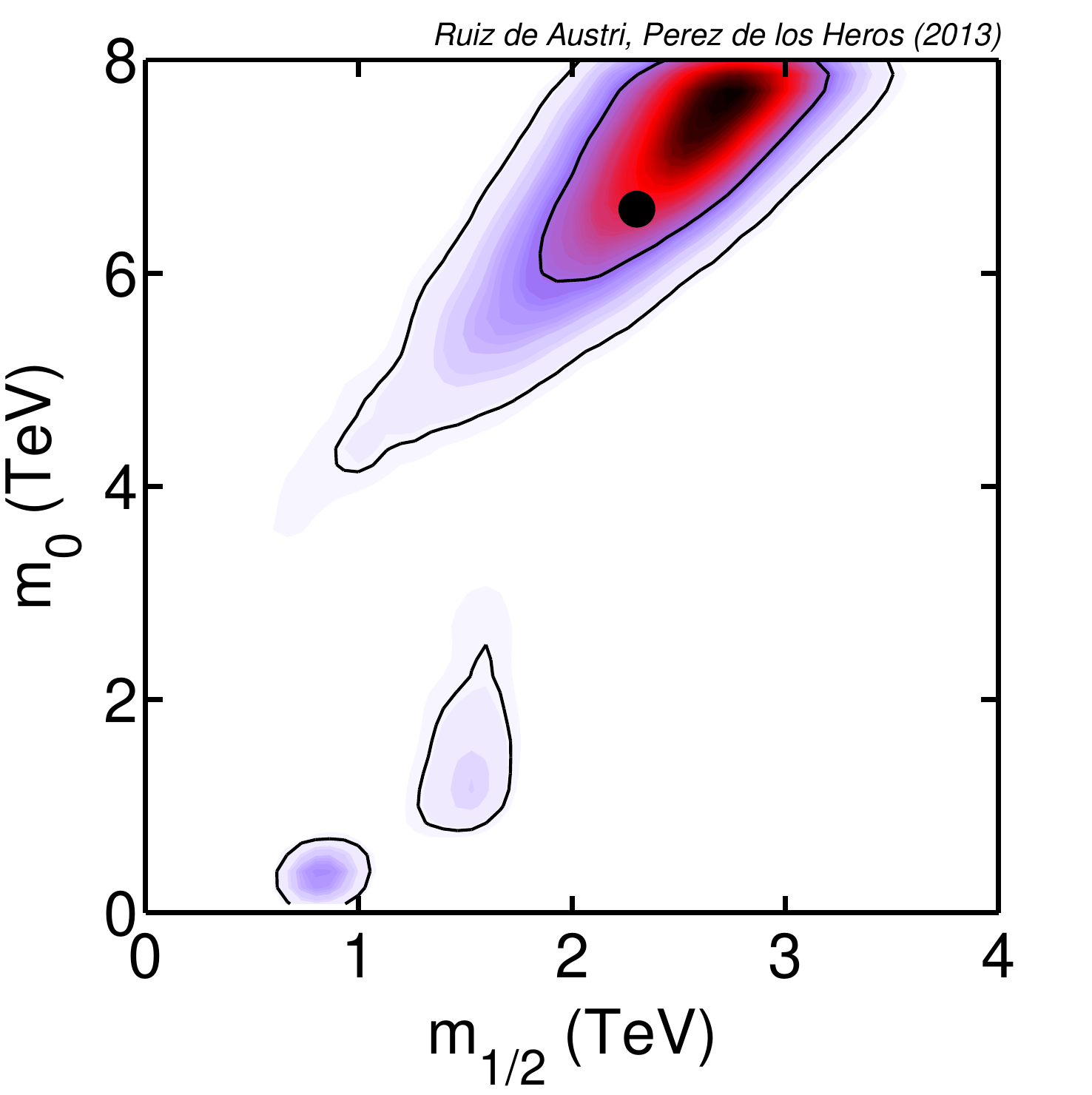}
\includegraphics[angle=0,width=.51\textwidth]{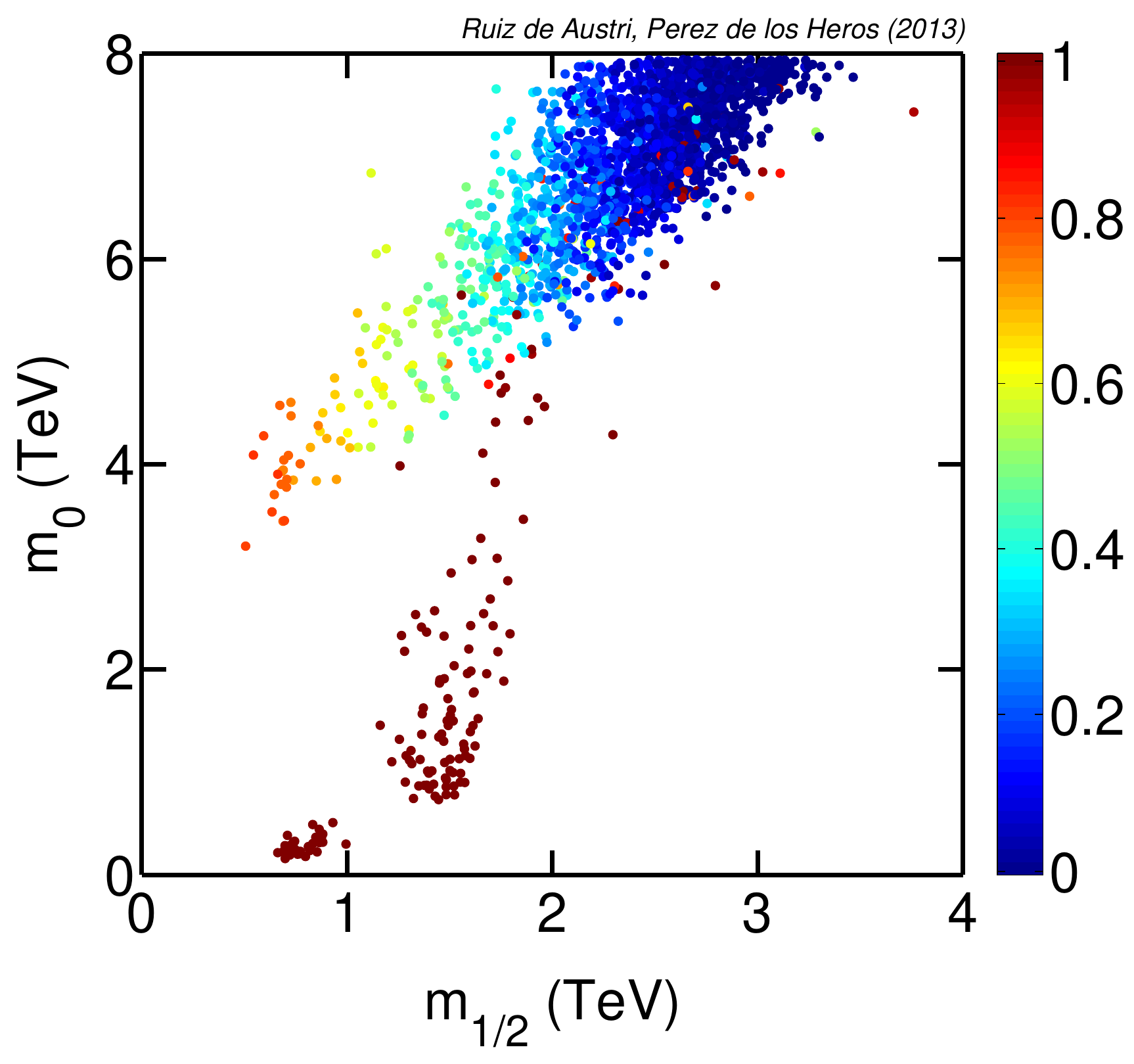}
\caption{\label{fig:mM} {\bf Left:} 2D marginalized posterior pdf of the ($m_{1/2}$,$m_{0}$) plane from the cMSSM
scan including particle physics and cosmological constraints from table~\ref{tab:exp_constraints}, excepting
dark matter detection constraints. 
The inner and outer contours enclose respective 68\% and 95\% joint regions. The filled circle indicates the posterior mean.
{\bf Right:} As in the left plot but showing the composition of the neutralino as gaugino fraction
in the color scale. The density of samples reflects the probability density.
}
\end{figure}

In this section we present the results of the different scans we performed and comment on the
conclusions that can be extracted on the effect of using LQCD or the experimental
determination of the nucleon matrix elements in interpreting experimental results.

\subsection{Results with no dark matter detection data} 

We begin by showing in Fig.~\ref{fig:mM} the impact of the current experimental constraints on the cMSSM 
from the inputs shown in table~\ref{tab:exp_constraints}, not including XENON100 and IceCube 
data. The results are shown in the ($m_{1/2}$,$m_{0}$) plane. The left panel displays the
marginalized posterior pdf including the 68\%, 95\% credible intervals whilst the right panel shows 
the gaugino fraction of the neutralino. 

Most of the posterior pdf lies in the high-mass region due to the Higgs mass measurement  
reported by ATLAS and CMS. The reason is that large radiative corrections are needed to
reconciliate theory and experiment and this can be achieved either with stops masses
$\gtrsim 3$ TeV and non-small $tan \beta$
to enhance the logarithmic corrections or/and through the maximal mixing scenario
where the stop mixing parameter $X_t = A_t - \mu \cot \beta$ satisfies
$X_t = \pm \sqrt{6} M_{SUSY}$ being $M_{SUSY}$ a certain average of the stop masses.
Whereas the later condition implies a certain degree of fine-tunning, the former is easy to
accommodate requiring large soft-mases \cite{Djouadi:2005gj}. 
Therefore it is expected that large gaugino and scalar masses are statistically favored. 
This region corresponds to the so called Focus-Point 
region \cite{Feng:1999mn} where scalar masses are multi-TeV while there is not 
much fine-tunning, thus, preserving the naturalness criterion
\footnote{The authors of \cite{Cabrera:2008tj,Cabrera:2009dm, Cabrera:2012vu} have shown how the 
fine-tunning penalization arises naturally in the Bayesian framework.} 
and where the neutralino is a Higgsino or a mixture of Higgsino-bino. It is precisely the Higgsino fraction which 
makes its self-annihilation very efficient to gauge bosons and top quarks. Besides, it coannihilates with the 
lightest charginos and the second lightest neutralino.
Actually for low or intermediate masses the annihilation is so efficient that its relic density can be
below the measured dark matter relic density. Thus, it needs a sizable bino fraction to be a viable 
dark matter candidate.
As long as the gaugino mass increases, the neutralino gets a larger Higgsino fraction until becoming
pure Higgsino. In this limit, the largest acceptable mass is $m_\neut \simeq 1$ TeV \cite{ArkaniHamed:2006mb}.
This is verified in the right panel of Figure 1, which shows the gaugino fraction of the neutralino, $g_f$. 
This fraction is defined as $g_f = |N_{11}|^2+|N_{12}|^2$, where the $N_{1i}$ represent 
the bino and wino component of the lightest neutralino respectively.
The other regions of the cMSSM where the required dark matter density is reproduced, namely, the 
stau-coannihilation  region where the neutralino LSP coannihilates with the lightest stau, 
and the A-funnel region where $2 m_\neut \simeq m_A$ and 
the annihilation goes via a resonance process to pairs of fermions
appear at the 95\% credible level and are statistically disfavored relatively to the Focus-Point 
region as they correspond to rather small soft-masses.

Since quite high supersymmetric masses are favored, the supersymmetric effects on g-2 and B-physics 
observables are negligible, thus, their role in the analysis is diluted.

\begin{figure}[t]
\includegraphics[angle=0,width=0.5\textwidth]{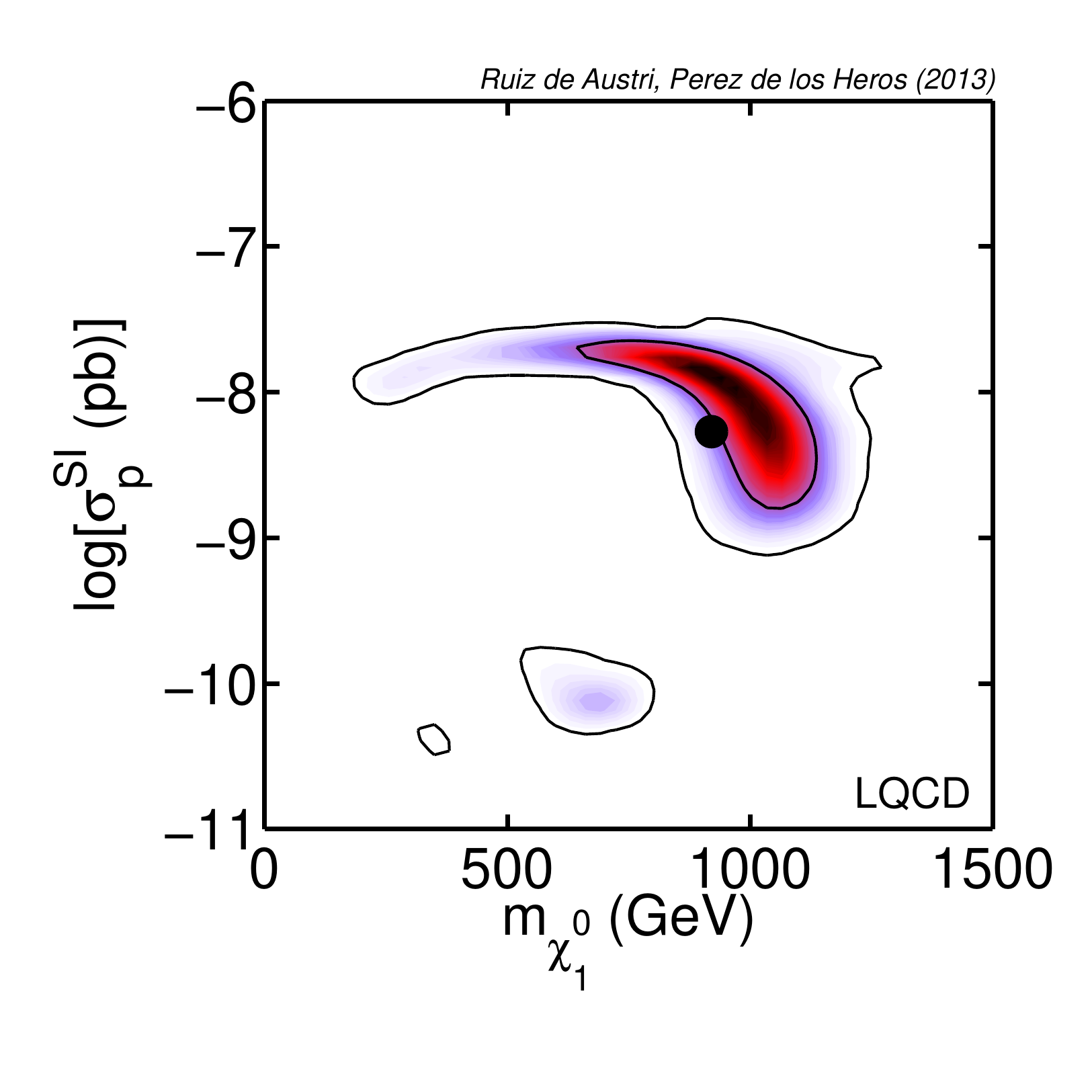}
\includegraphics[angle=0,width=0.5\textwidth]{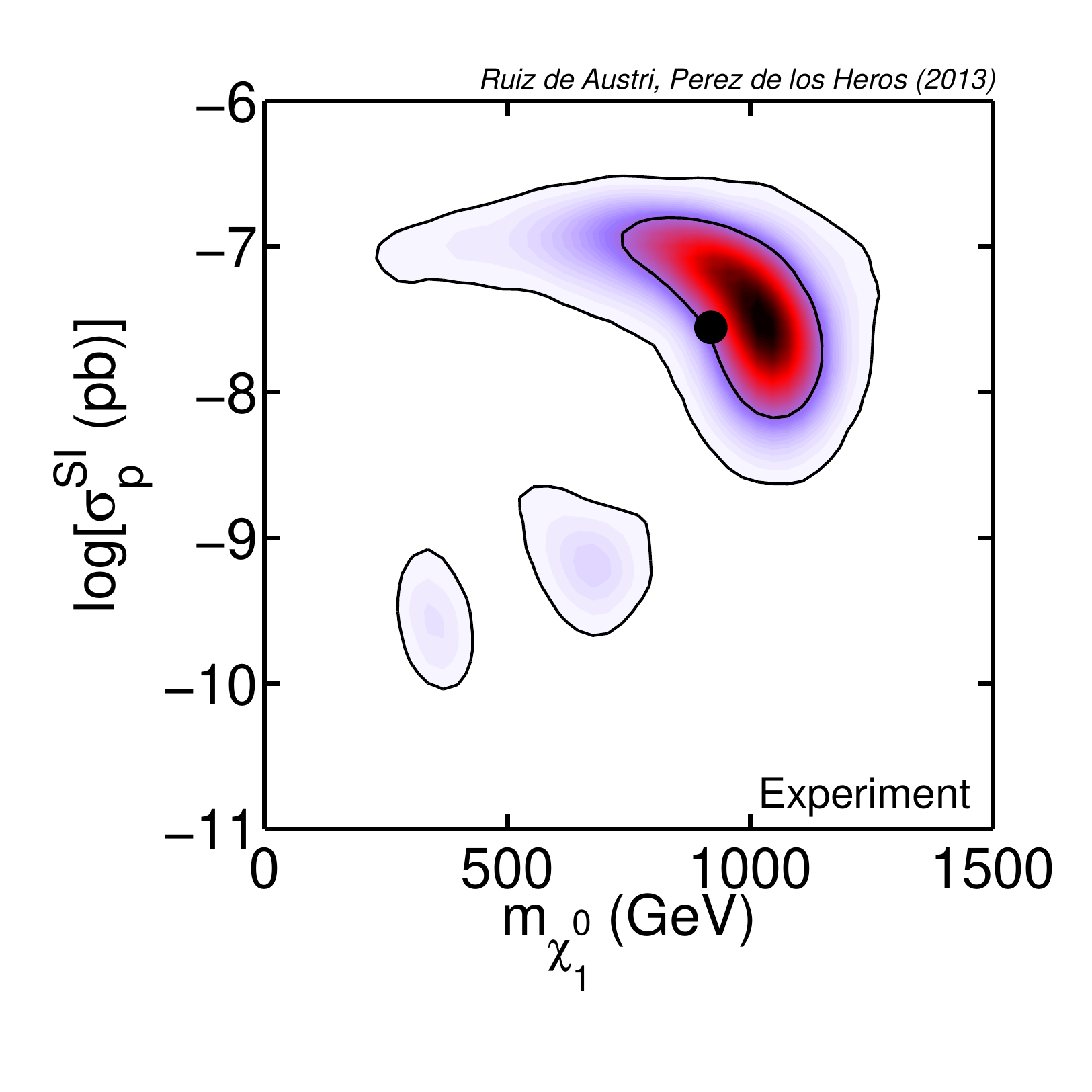} \\ 
\vspace{0.5cm}
\includegraphics[angle=0,width=.5\textwidth]{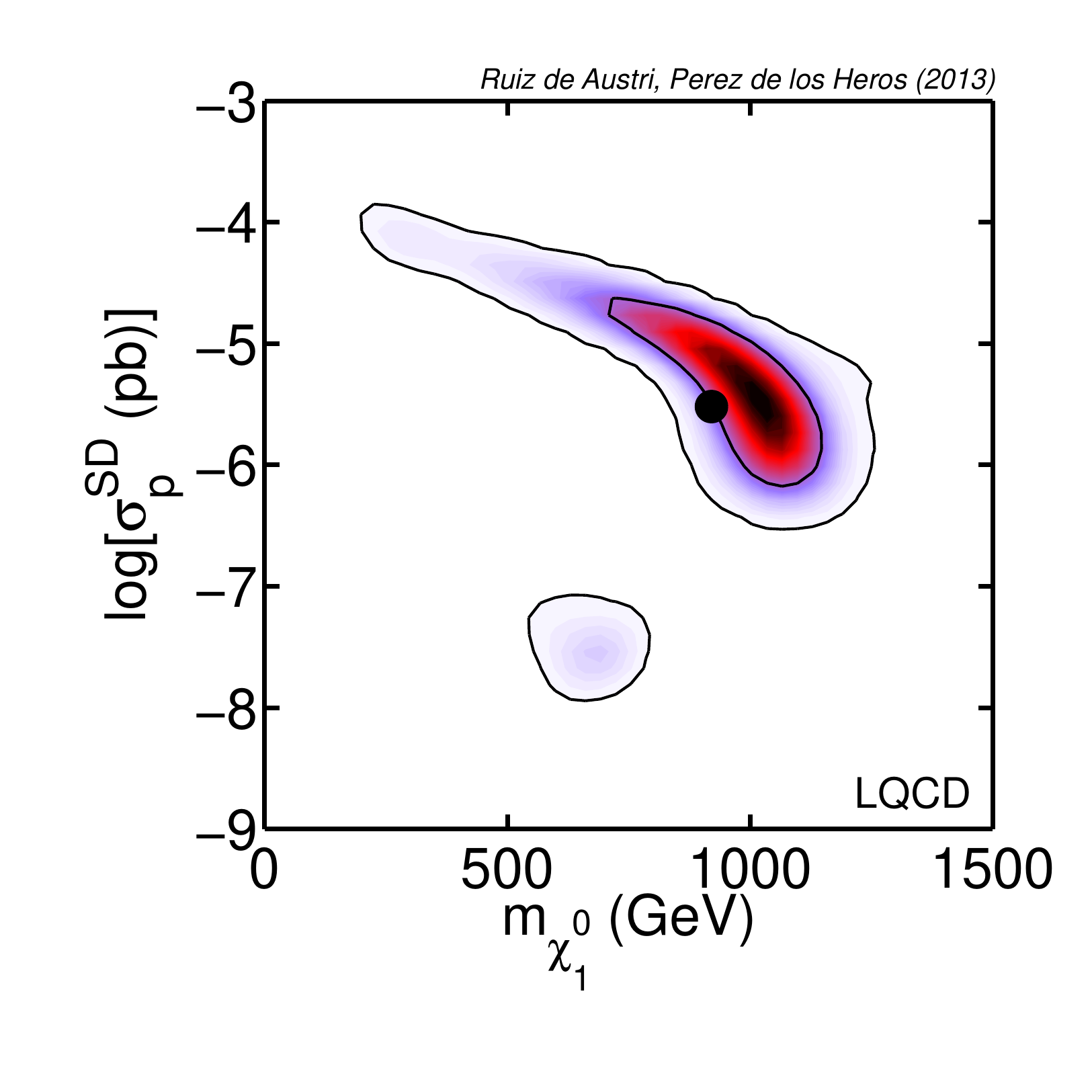}
\includegraphics[angle=0,width=.5\textwidth]{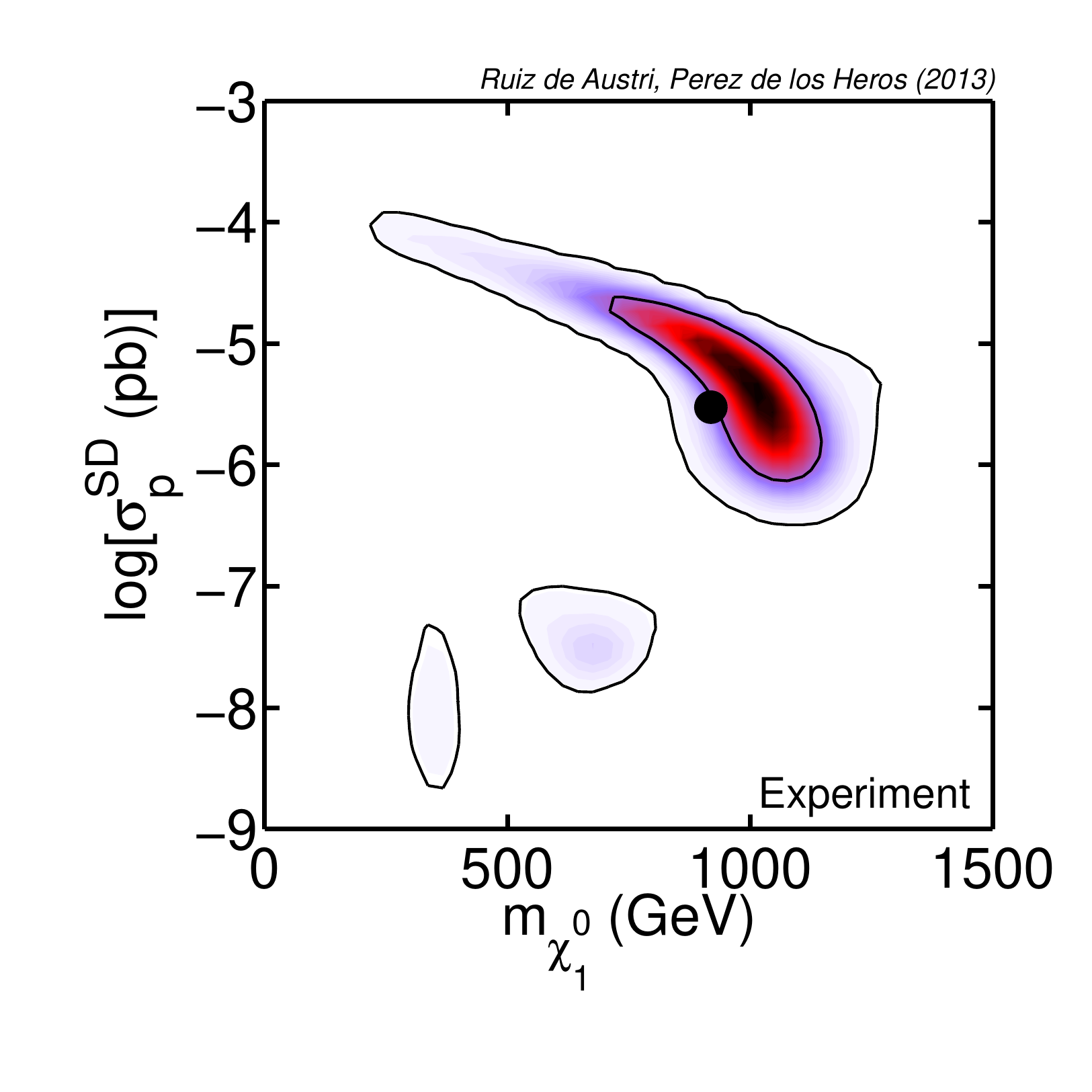}
\caption{
\label{fig:sigma_no}
{\bf Upper panel:} 2D marginalized posterior pdfs of the ($m_\neut, \sigmaSI$) plane obtained using the 
values of the hadronic structure functions from LQCD  (left) and experimental calculations (right) with all 
the particle physics and cosmological constraints in table~\ref{tab:exp_constraints} included, 
excepting dark matter detection constraints.
{\bf Lower panel:} Same as above but for the ($m_\neut, \sigmaSD$) plane.}
\end{figure}

Fig.~\ref{fig:sigma_no}, shows the posterior pdfs in the ($m_\neut, \sigmaSI$) plane 
(upper panels) and in the ($m_\neut, \sigmaSD$) plane (lower panels). 
The left panels use the LQCD estimation of the nucleon matrix elements, whereas the 
right ones the experimental measurement assuming that $\sigma_{\pi N} = 74$ MeV as outlined in 
Section \ref{sec:intro}.

Following the discussion above, it is clear that the bulk of the posterior pdf lies in the Focus-Point region
which covers neutralino masses from a few hundred GeV to $\sim 1$ TeV.
In this region the spin-independent cross section, shown in the top panels, is large
because the neutralino is a mixed bino-Higgsino state and the dominant diagrams entering in the 
process are mediated by a Higgs H/h which scale as $\propto |N_{11}|^2 |N_{14/13}|^2$, where 
$N_{14/13}$ represent the Higgsino composition of the lightest neutralino.
As long as the gaugino mass increases the neutralino eventually becomes a pure Higgsino 
and the sensitivity is lost.
In the stau-coannihilation and the A-funnel regions, the neutralino is bino-like and therefore the spin-independent 
cross section is suppressed. The A-funnel region has a larger spin-independent cross section because $tan \beta$ 
is typically larger and the heavy Higgs contribution is enhanced with respect to the one in the stau-coannihilation region.

The effect of using different determinations of the nucleon matrix elements in
the spin-independent cross section is dramatic, as can be seen by comparing the left and right top panels.
The posterior pdf is shifted by almost a factor 10 due to the large differences in the estimation
of the strangeness content of the nucleon $f_{T_s}$ , which is the dominant contribution in the evaluation of 
the spin-independent cross section, between LQCD and the experimental approaches. 
This is precisely the difference since $f_{T_s}$ acts as a proportionality factor in the coupling of 
the neutralino to the nucleon, as shown in Eq.~\ref{eq:f_t}. 
Therefore it is expected that results from XENON100 on the spin-independent cross section disfavors a 
larger portion of the Focus-Point when using the experimental values of $f_{T_s}$, thus having a bigger impact on the  
($m_{1/2}$,$m_{0}$) plane. We will come back to this point later.

Let us now discuss the spin-dependent cross section which is shown in the lower panels 
of Fig. ~\ref{fig:sigma_no}.
Due to the LHC bounds on squark masses which currently are constrained to be above $\sim 1$ TeV, 
this is largely governed by Z-boson exchange and is sensitive to the Higgsino 
asymmetry $\sigmaSD \propto (|N_{13}|^2 -|N_{14}|^2)^2$ since the 
bino is a SU(2) singlet and it does not couple to the Z-boson. Therefore the spin-dependent cross
section is significantly increased in the mixed bino-Higgsino neutralino scenario
while being suppressed for a pure Higgsino-like neutralino.
This is precisely what is shown in the plots. 

Comparing the upper and lower panels one can see that in contrast  to the spin-independent counterpart, 
the spin-dependent cross section is remarkably stable to the choice of the nucleon matrix elements type of 
determination. 
This is because, for the latter, the dominant contribution of the nucleon matrix elements 
come from the up/down quark flavors which are consistent 
at the 1-$\sigma$ level within LQCD and the experimental approaches.

\begin{figure}[t]
\includegraphics[angle=0,width=.5\textwidth]{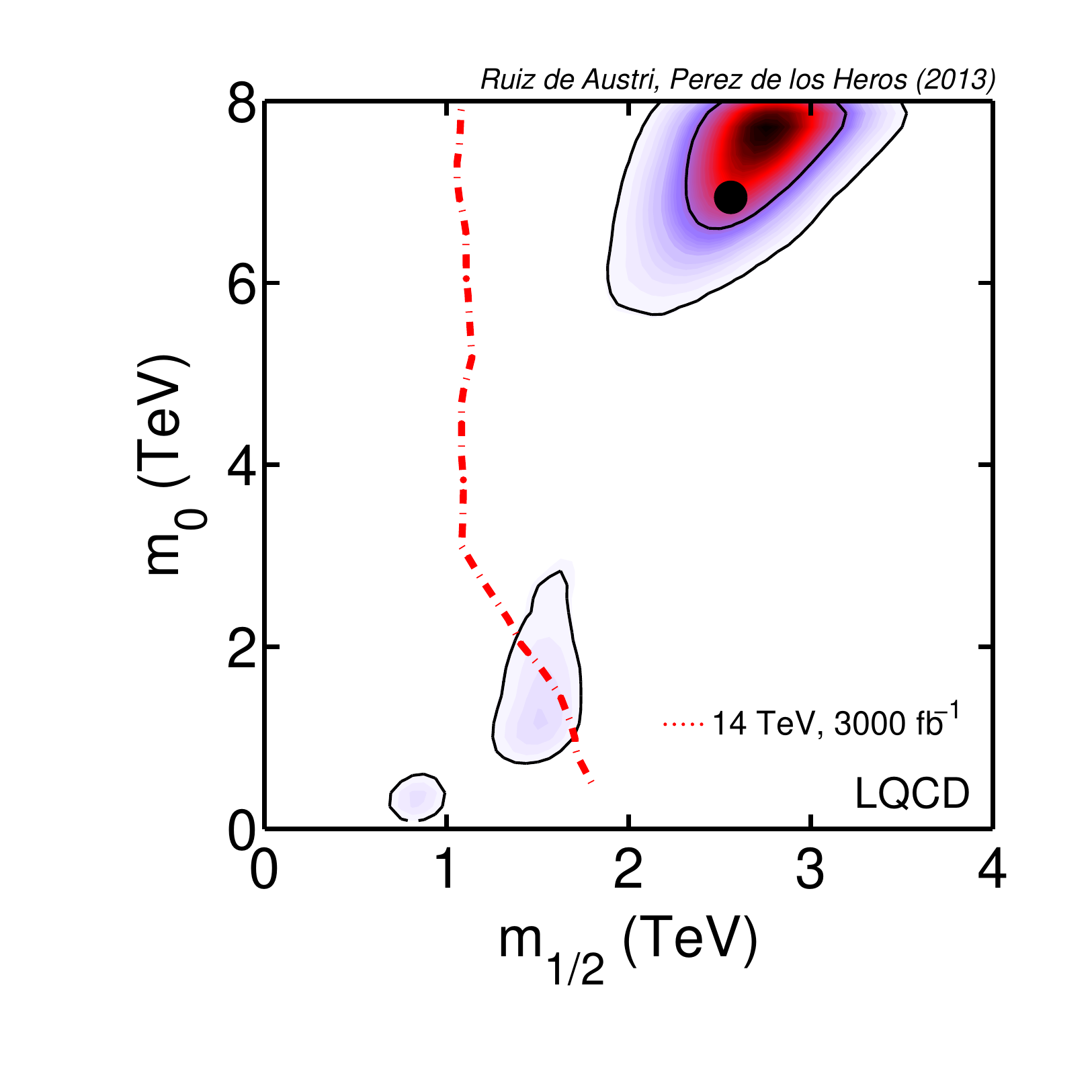}
\includegraphics[angle=0,width=.5\textwidth]{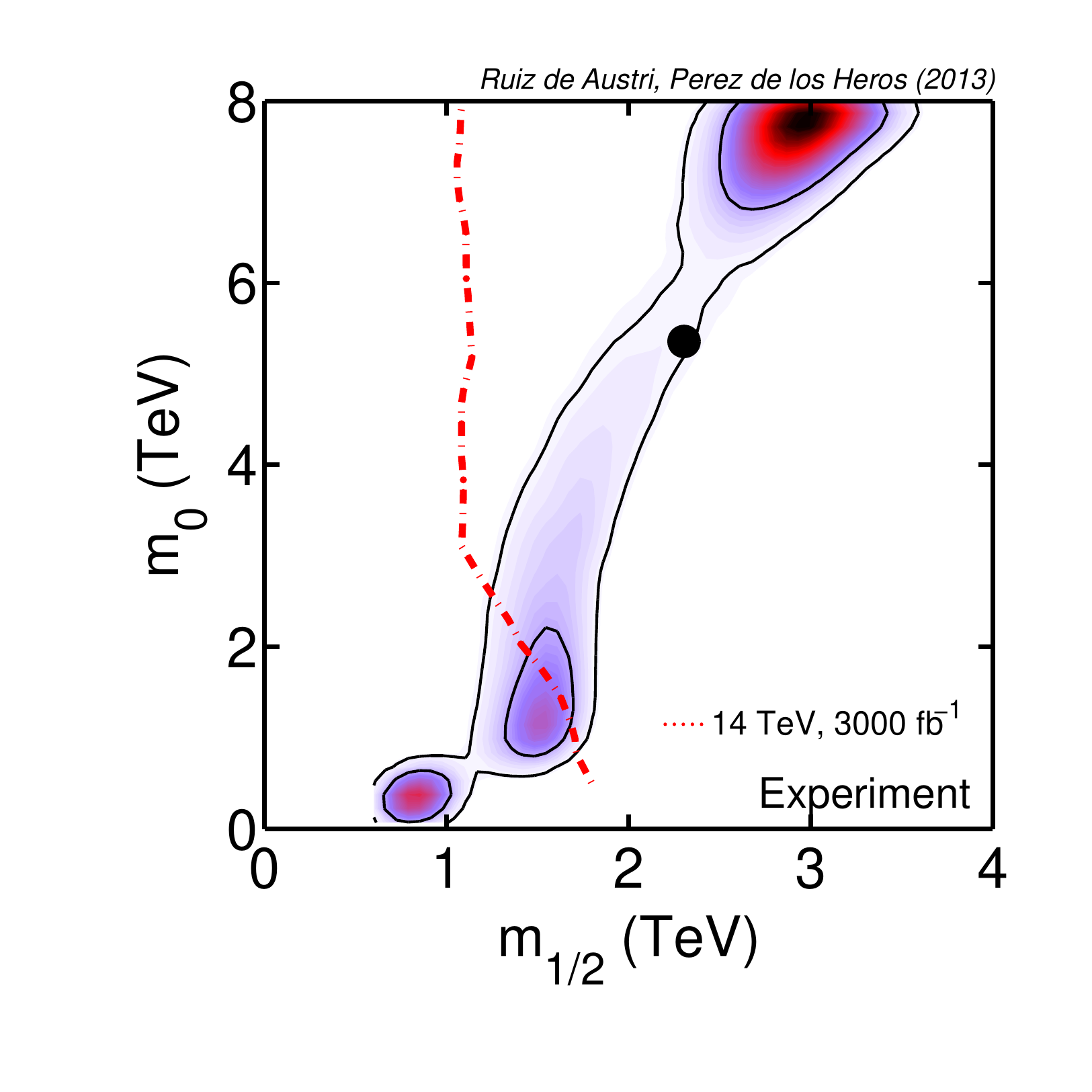} \\
\vspace{0.5cm}
\includegraphics[angle=0,width=.5\textwidth]{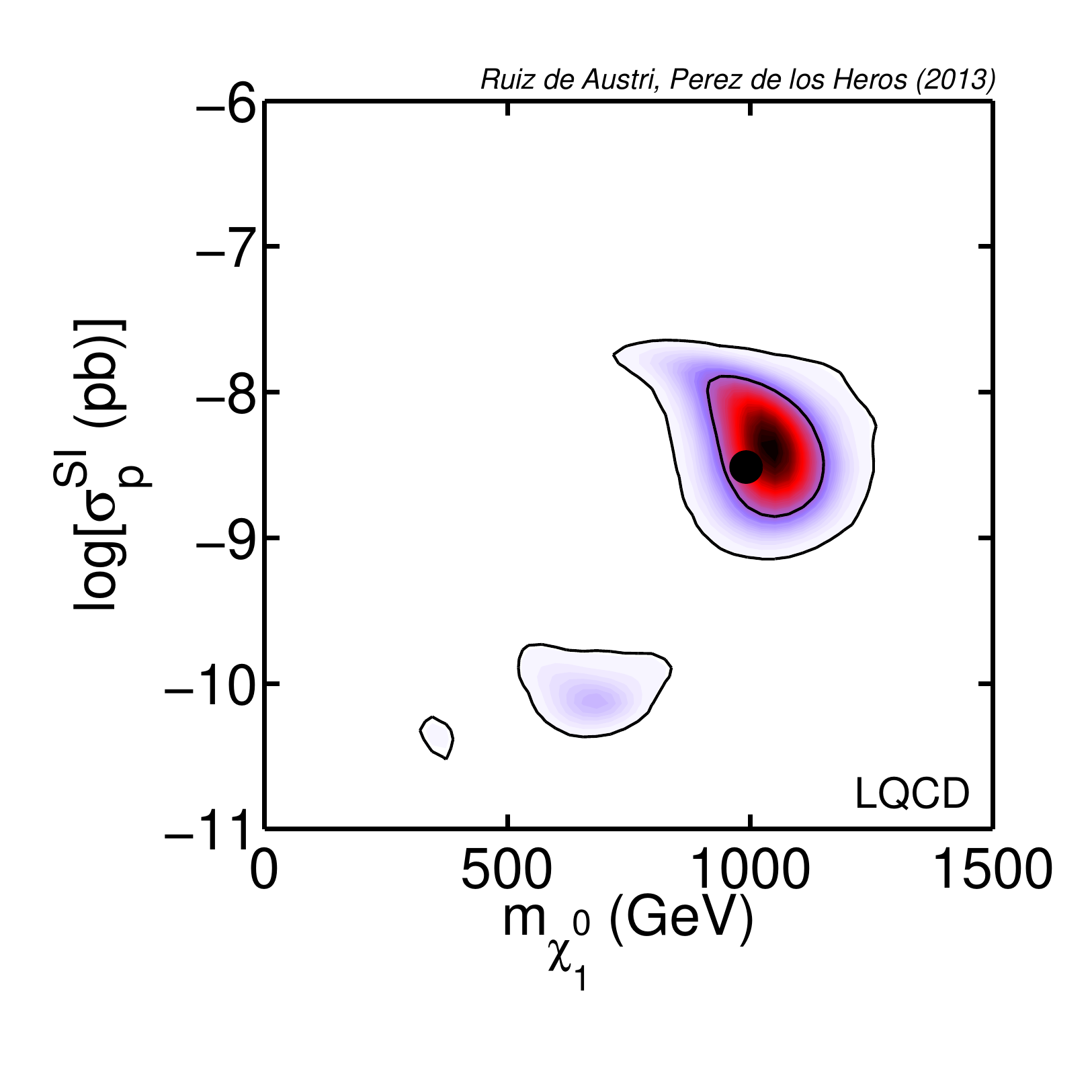}
\includegraphics[angle=0,width=.5\textwidth]{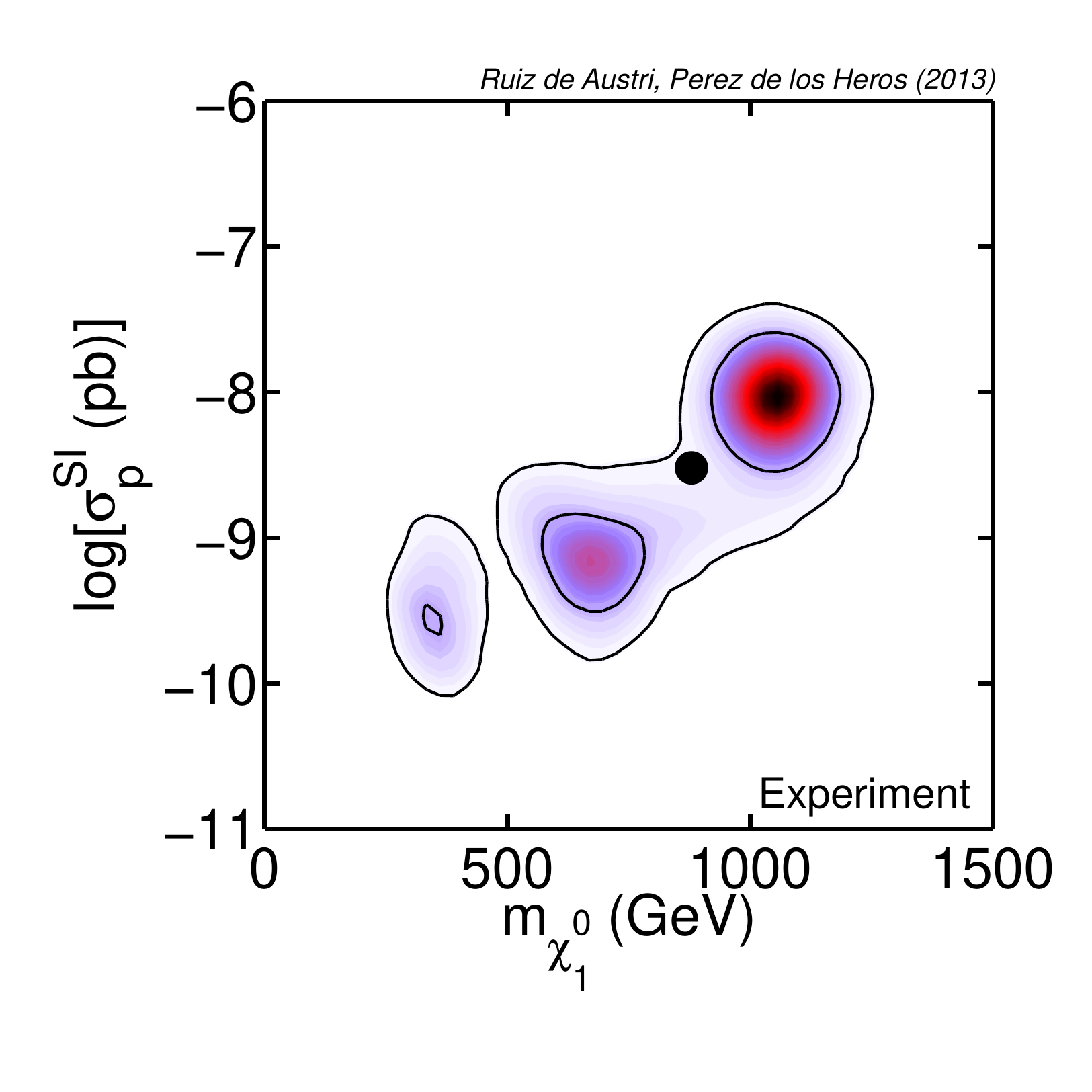}
\caption{\label{fig:sigma_Xe}
{\bf Upper panel:} 2D marginalized posterior pdfs of the ($m_{1/2}$,$m_{0}$)  plane  
from the cMSSM scan including particle physics and  cosmological constraints from table~\ref{tab:exp_constraints}
as well as XENON100 constraints, 
obtained using the values of the hadronic structure functions from LQCD (left) and experimental calculations (right). 
The dot-dashed/red line shows the projected ultimate LHC reach in the high-luminosity phase with an energy of  
$14$ TeV  and an integrated luminosity of  3000 fb$^{-1}$ (from \cite{HL_LHC}).
{\bf Lower panel:} Same as above but for the ($m_\neut, \sigmaSI$) plane. }
\end{figure}

\subsection{Results including dark matter direct detection data} 

Next, we focus on the impact of XENON100 data on the cMSSM which we show in Fig. ~\ref{fig:sigma_Xe}.  
The upper panels show the ($m_{1/2}$, $m_{0}$) plane when limits from  
XENON100 are included. The lower panels show the ($m_\neut, \sigmaSI$) plane. Left panels use the LQCD 
determination of the nucleon matrix elements whereas the right ones the experimental one. Clearly XENON100 data 
disfavors the Focus-Point region, as expected from the discussion above, since the neutralinos are mixed 
bino-Higgsinos with masses of $\cal{O}$(100 GeV). 
The role of the XENON100 data on the ($m_{1/2}$, $m_{0}$) plane changes dramatically when  
either the LQCD or the experimental approaches are used to extract the strangeness content 
of the nucleon. In the latter case a larger portion of the Focus-Point region is disfavored and therefore a 
sizable fraction of the posterior pdf is displaced to the stau-coannihilation and A-funnel regions which now are favored 
at the 68 \% credible level. 

As an example, we display the projected ultimate LHC reach in the high-luminosity phase with an 
energy of  $14$ TeV  and an integrated luminosity of  3000 fb$^{-1}$ \cite{HL_LHC}. 
One can conclude  that, regarding LHC searches, there are much better detection 
prospects in the case of assuming  higher values of the $f_T$ coefficients, 
represented by the experimentally obtained values in our study. 

\begin{figure}[t]
\includegraphics[angle=0,width=.5\textwidth]{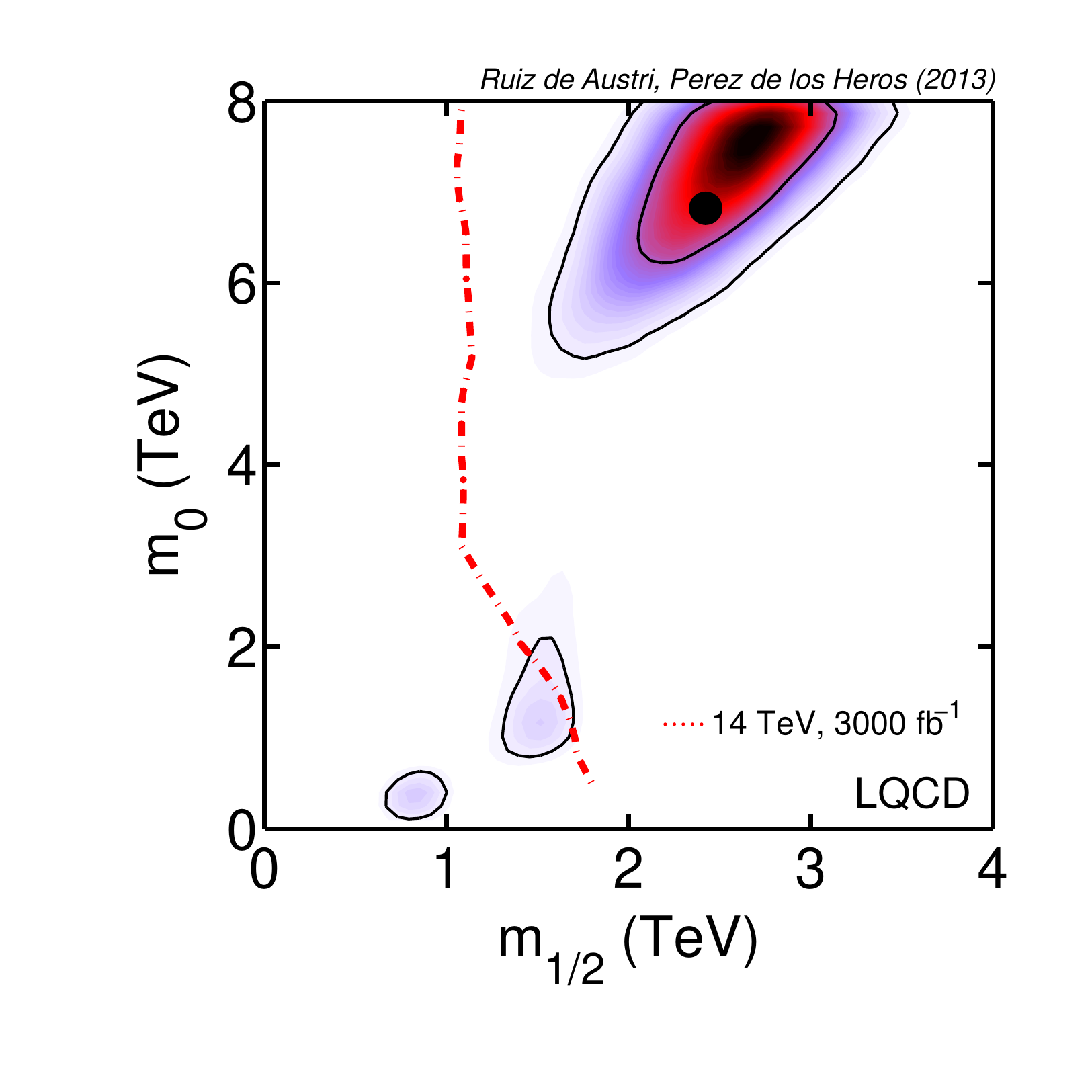}
\includegraphics[angle=0,width=.5\textwidth]{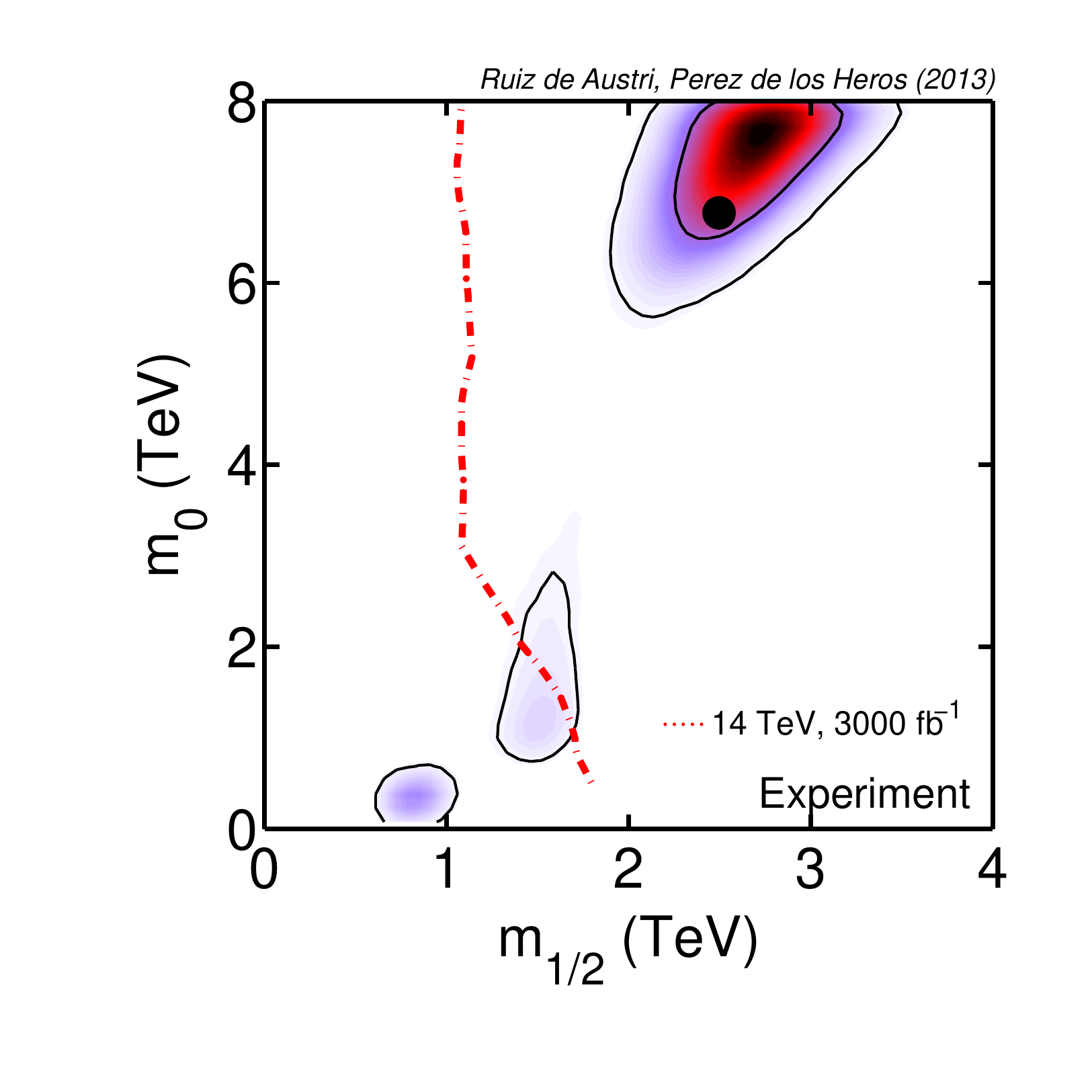} \\
\vspace{0.5cm}
\includegraphics[angle=0,width=.5\textwidth]{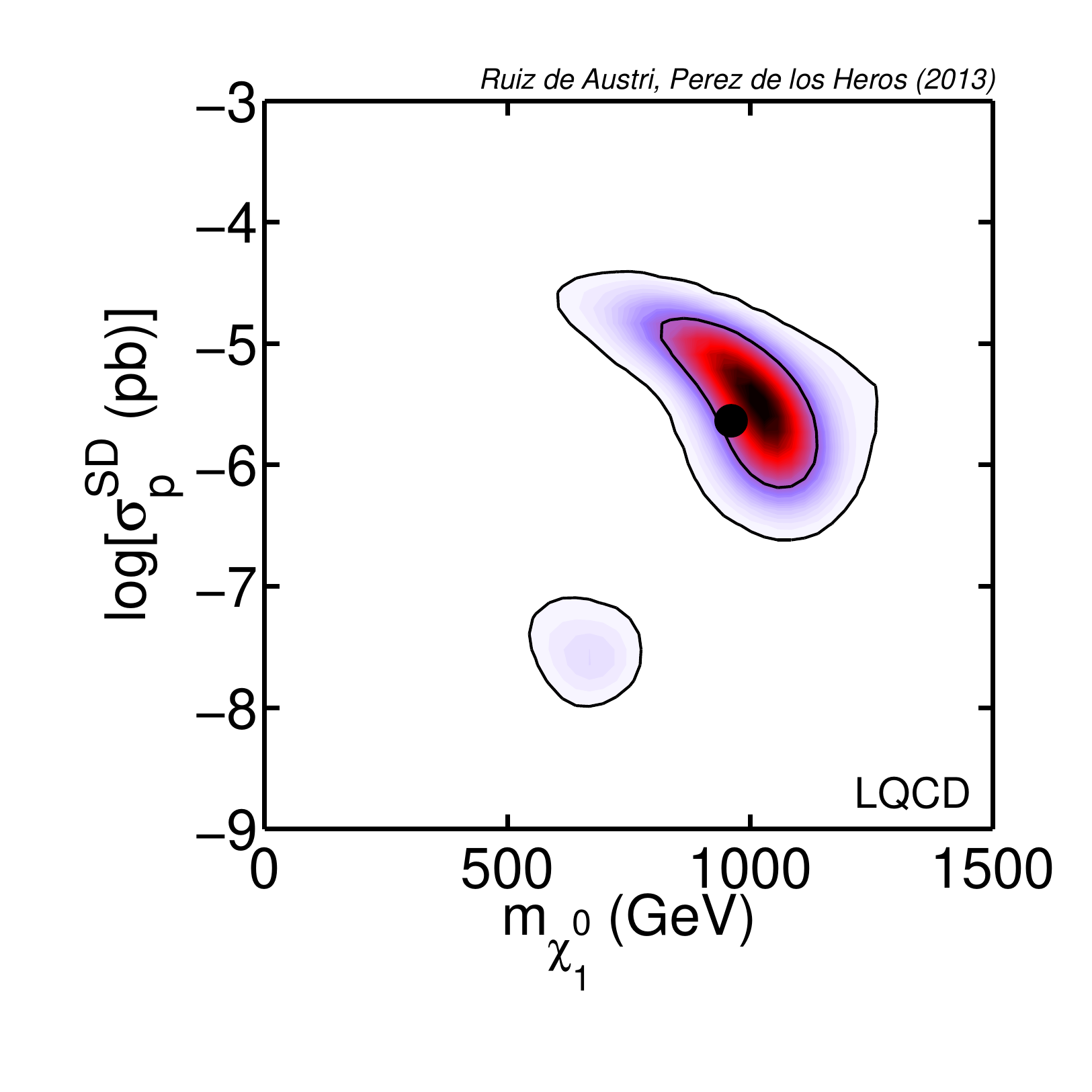}
\includegraphics[angle=0,width=.5\textwidth]{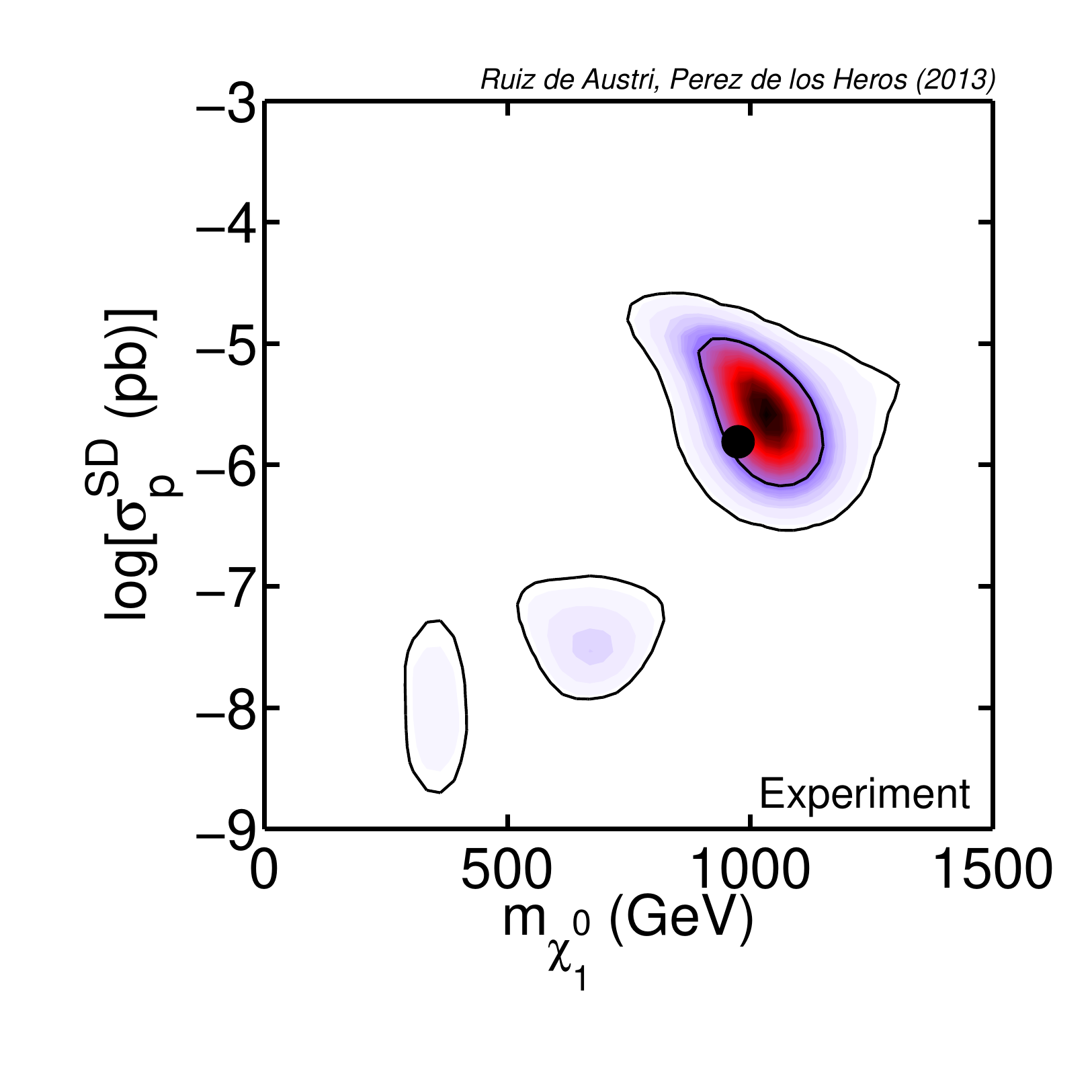}
\caption{\label{fig:sigma_IC}
{\bf Upper panel:} 2D marginalized posterior pdfs of the ($m_{1/2}$,$m_{0}$)  plane  
from the cMSSM scan including  particle physics and cosmological constraints from table~\ref{tab:exp_constraints}
and IceCube86 constrains, but excepting XENON100 constraints, obtained using the values of the 
hadronic structure functions from LQCD (left) and experimental calculations (right). The dot-dashed/red line shows 
the projected ultimate LHC reach in the high-luminosity phase with an energy of  $14$ TeV  and an integrated luminosity of  
3000 fb$^{-1}$ (from \cite{HL_LHC}).
{\bf Lower panel:} Same as above but for the ($m_\neut, \sigmaSD$) plane. }
\end{figure}

The ($m_\neut, \sigmaSI$) plane shows that XENON100 data have a
significant softer impact when LQCD nucleon form factors are employed. 
In this case (left-panel) neutralinos of the bino-Higgsinos mix type are still favored in the Focus-Point region at the 
95 \% credible level, whereas when applying experimental data, in essence, only the pure Higgsino 
scenario remains in the Focus-Point region.
The connected bimodal type of shape shown by the posterior pdf is an effect of the mentioned 
displacement of the posterior towards the A-funnel funnel region where lower spin-independent cross sections are
favored. The remaining probability island which exists at the 68 \% credible level corresponds 
to the stau-coannihilation region.



\subsection{Results including dark matter indirect detection data} 

Fig. \ref{fig:sigma_IC} shows the impact of IceCube86 data only (i.e. non including XENON100 data) in the
($m_{1/2}$, $m_0$) plane on the upper panels and ($m_\neut, \sigmaSD$) on the lower
panels.

As pointed out previously, the spin-independent and spin-dependent cross sections are correlated through the Higgsino fraction 
of the neutralino. Actually as long as the Higssino fraction is larger than $\cal{O}$(10\%), cross sections 
which are at reach to both XENON100 and IceCubeC86 can be achieved. Therefore one expects 
that the Focus-Point  region be also probed by IceCube86 as it is done by XENON100. This is what can be  
observed in the upper panels of Fig. \ref{fig:sigma_IC}. Of course, in points where the neutralino 
becomes Higgsino dominated, the spin-dependent cross section drops and the sensitivity is lost. 
In comparison with the SI counterpart, the posterior pdf  is remarkably stable to the choice 
of the method to determine the nucleon matrix elements as argued above (only a small fraction of the posterior 
pdf is displaced to both the A-funnel and stau-coannihilation regions that now remain disconnected though). 
Therefore the bulk of the posterior pdf lies well in the Focus-Point for both choices. 
In this case one can conclude that the chance to probe the model at the LHC is small in 
either approach.

The lower panels of Fig.~\ref{fig:sigma_IC} show the same behavior: the allowed region in the 
($m_\neut, \sigmaSD$) plane hardly changes in the two scenarios considered. The 68~\% credible level which 
corresponds to Higgsino-like neutralinos is qualitatively similar between the LQCD and the experimental approaches,  
and just the tails differ, allowing lighter neutralinos in the former case, with cross sections above $\sim 10^{-6}$ pb. 



 

\section{Conclusions}\label{sec:concl}

In this work we have investigated the impact of using different values of the nucleon matrix elements in inferences 
of the cMSSM parameter space, though the conclusions can be extrapolated qualitatively to any other SUSY model 
without loosing generality. Among the wide range of values of the nucleon matrix elements found in the literature 
obtained by different groups, we have chosen typical extreme values of the range to illustrate the impact they can 
have on the interpretation of dark matter searches. We have used estimations from recent LQCD calculations 
as well as experimental results, using these different values of the nucleon matrix elements as inputs to 
different scans over the cMSSM parameter space.

We have shown that the role of dark matter experiments sensitive to spin-independent cross sections, like 
XENON100, is strongly affected by the large differences in the determination of the strangeness content of the 
nucleon. The reason is that spin-independent cross sections can vary up a factor $\sim$ 10 depending on which 
input for the nucleon matrix elements is used. The posterior pdf of a given model is displaced rougly by that 
factor.  The immediate result is that a larger portion of the Focus-Point region is disfavored by current direct 
dark matter searches when using nucleon data that favor a larger strangeness content of the nucleon. Note that the 
most disfavored regions from the Higgs mass constraint from the LHC (not including XENON100 data), 
as the stau-coannihilation and A-funnel regions, get a high statistical weight when using a high value 
of the nucleon matrix elements, represented in our study by the experimental value. This has a strong impact for 
the interpretation of SUSY searches at the LHC. The accessible region of parameter space changes radically whether 
low or high values of the matrix elements are used. However, if future ton-scale direct dark matter 
experiments can probe, as expected, spin-independent cross section levels of about $10^{-11}$ pb, 
the dependence on the matrix elements will be alleviated, since at that level that dependence becomes 
negligible and, at least in the case of the cMSSM, practically all the currently allowed parameter space can be 
probed.

The conclusion is more favorable for experiments sensitive to the spin-dependent cross section, like neutrino 
telescopes. They are practically not affected by the choice of values of the nuclear axial-vector matrix elements 
which drive the spin-dependent neutralino-nucleon cross section.
The reason is that even if there is a large difference in the strangeness content of the nucleon for spin-dependent 
interactions, light quarks play here a prominent role, and the determination of their content in the nucleon from 
LQCD calculations or experimental results is consistent at 1-$\sigma$ level. To illustrate this we have applied 
constraints from IceCube in our cMSSM scans assuming a background--only scenario with the 86 strings configuration, 
inspired by their recent null result on searches for an excess muon flux from neutralino annihilations in 
the Sun. In view of the results shown in Fig.~\ref{fig:sigma_IC} it can be stated that current limits from neutrino 
telescopes on the spin-dependent neutralino-nucleon cross section are robust in what concerns the choice of nucleon  
matrix elements, and these quantities should not be a concern in interpreting neutrino telescope results.

\acknowledgments
 We thank the Kavli Institute for Theoretical Physics at UCSB and organizers of the Hunting for 
Dark Matter programme for their hospitality during the preparation of this manuscript. 
This research was supported in part by the National Science Foundation under Grant 
No. NSF PHY11-25915. 
R. RdA, is supported by the Ram\'on y Cajal program of the Spanish MICINN 
and also thanks the support of the 
Spanish MICINN's Consolider-Ingenio 2010 Programme under the grant MULTIDARK 
CSD2209-00064 and the Invisibles European ITN project (FP7-PEOPLE-2011-ITN, 
PITN-GA-2011-289442-INVISIBLES). 
The use of IFT-UAM High Performance Computing Service is gratefully 
acknowledged.


\end{document}